\def \bc {\begin{center}}
\def \ec {\end{center}}

\def \bfr {\begin{flushright}}
\def \efr {\end{flushright}}

\def \v {\vskip}

\def \ba {\begin{array}}
\def \ea {\end{array}}

\def \bea {\begin{eqnarray}}
\def \eea {\end{eqnarray}}

\def \be {\begin{equation}}
\def \ee {\end{equation}}
\def \p {\partial}

\def \d {\hbox{d}\,}
\def \s {\hbox{s}}

\def \e {\hbox{e}}

\documentstyle[12pt,a4]{article}
     
\begin{document}
%

\centerline{}
\thispagestyle{empty}
\vskip 3 true cm 

\begin{center} 
{\bf 2D DILATON GRAVITY MADE COMPACT}
\vskip5mm 
 Miguel Navarro\footnote{http://www.ugr.es/
$\widetilde{\ }$mnavarro; mnavarro@ugr.es} 

\end{center}
\normalsize
\v2mm

{\it Instituto de Matem\'aticas y F\'\i sica Fundamental, 
        CSIC. Serrano 113-123, 28006 Madrid, Spain.}             

\centerline{and}
\vskip 0.5 true cm 
{\it Instituto Carlos I de F\'\i sica Te\'orica y Computacional,
        Facultad  de  Ciencias, Universidad de Granada. 
        Campus de Fuentenueva, 18002, Granada, Spain. }

\vskip10mm

\begin{center}
                        {\bf Abstract}
\end{center}

\footnotesize 

We show that the equations of motion of two-dimensional 
dilaton gravity conformally coupled to a scalar 
field can be reduced to a single non-linear second-order partial differential 
equation when the coordinates are chosen to coincide with the two 
scalar fields, the matter field $f$ and the dilaton $\phi$, 
which are present in the theory. This result may help 
solve and understand two- and higher-dimensional classical and quantum 
gravity. 

\normalsize

\vskip 3mm 

\v2mm
\noindent Keywords: 2D gravity, differential equations, 
solvability. 
\newpage{}
\setcounter{page}{1}
Low-dimensional models of gravity are receiving a great 
deal of attention lately as they can provide  
insight into the classical as well as the 
quantum theories of more realistic (higher-dimensional) 
theories of gravity. 
Prominent among these low-dimensional general covariant 
dynamical systems are the 2D dilaton models of gravity, 
the general action of which can be written in the form 

\be {S}_{{GDG}}= S_V - S_M\label{GDG}\ee 
where
\be S_V=\int d^2x\sqrt{-{g}}
\left({R}{\phi} + {V} ({\phi})\right) \label{matterlessGDG} \ee 
and $S_M$ is a gravity-matter interaction term which 
may be more general but we shall take in the present letter 
to be of the form 

\be S_M=\frac12\int d^2x\sqrt{-{g}}\,\Omega(\nabla f)^2\ee
 with $\Omega$ and unspecified function of the dilaton field 
$\phi$, $\Omega=\Omega(\phi)$. A direct connection can be 
made with physical reality by noting that, for instance,   
spherically symmetric Einstein-Hilbert gravity 
minimally coupled to a massless scalar field coincides with  
the model with $V=2/\sqrt{\phi}$ and 
$\Omega=G\phi$, with $G$ the Newton constant.  

Unfortunately, and notwithstanding their relative simplicity, 
most of these models, in particular spherically symmetric 
Einstein gravity, have eluded their being analitically solved at 
 the classical level, let alone the quantum one. 
Actually, solving these model is relatively easy 
when no matter is present, as they are highly symmetric 
\cite{[symmatter]}.  However, the introduction of matter 
fields breaks a great deal of these 
symmetries (although some are preserved) 
and makes solving them a much more difficult task  
\cite{[symmatter],[gensym]}.  
In fact, when coupled to conformal matter, they can 
be solved only for $V(\phi)=4\lambda^2 \e^{\beta\phi}$ 
with constant $\lambda$ and $\beta$  (the string-inspired 
and the exponential models) \cite{[CGHS],[conformal]} 
and for $V(\phi)=4\lambda^2\phi$ 
(the Jackiw-Teitelboim model) \cite{[JTmodel]}. 

For arbitrary $V$, these models have been solved only 
for chiral matter. In this case, the (generalized 
Vaidya) solution is given by \cite{[symmatter]}

\be \d\s^2= 2\d\phi \d u + (2M(u)-J(\phi)) (\d u)^2\label{Vaidya}\ee 
where  $\d J/d\phi=V$ and 

\be M(u)=\int^u \d\tilde{u} T_{uu}(\tilde{u})\label{Mu}\ee
with $T_{uu}$ the only non-null component of the energy-momentum 
tensor.  

Beyond these very particular cases,  
we have to resort, even at the classical level,  
to approximate solutions, such as the ones provided by 
numerical methods (see, for instance Ref. \cite{[Choptuik]} and references 
therein), perturbative methods (see, for instance 
Ref. \cite{[Mikovic]} and references therein) and so on. 
It is clear that this situation precludes our gaining
full benefit from these models in seeking 
insight into quantum gravity. 
This obstruction is even more grave because of the fact 
that some of the problems we must  
overcome before arriving at a quantum theory 
of gravity can be traced back to our failure 
to reach a full understanding of the classical theory. 
The problem of time, for instance, 
already exists at the classical level, 
 even though in this case it can be swept under  
the carpet \cite{[time]}. 
Also, how and what we can observe remains 
largely a mistery even in the classical theory 
\cite{[observable]}. In general, 
there is a good case for arguing that  
the current approach to observations in the classical theory 
of gravity is, at the very least, rudimentary, and perhaps  
badly conceived and altogether 
inadequate to be extrapolated to the quantum theory. 
For instance, in any theory which aspires    
to be truely fundamental, 
the space-time manifold cannot be taken 
as a given primary concept but rather  
it should be derived from more basic principles. 

The analysis in the present letter may help  
solve (and improve our understanding)  
of the classical two- and higher-dimensional 
models of gravity and help devise the quantum theories. 

The Euler-Lagrange equations of motion of the 
models in Eq. (\ref{GDG}) can be brought to the form: 

\bea 
R+V'(\phi)&=&T_\phi\label{eqofmotR}\\
\nabla_\mu\nabla_\nu\phi &=& 
\frac12g_{\mu\nu}V -T_{\mu\nu}\label{eqofmotphi}\\
\nabla_\mu\left(\Omega \nabla^\mu f\right)&=&0\label{eqofmotf}\eea 
where 

\be T_\phi=\frac{\Omega'(\phi)}2(\nabla f)^2\ee
and

\be T_{\mu\nu}=\frac{\Omega}2\left\{\nabla_\mu f\nabla_\nu f
-\frac12g_{\mu\nu}(\nabla f)^2\right\}\ee 

As shown in Ref. \cite{[symmatter]}, Eq. (\ref{eqofmotR}) 
can be deduced from Eq. (\ref{eqofmotphi}) by using basic 
properties of the covariant derivatives $\nabla_\mu$ and the 
curvature tensors in two dimensions. Therefore, 
we can concentrate on Eq. (\ref{eqofmotphi}) and 
Eq. (\ref{eqofmotf}). 

For chiral matter, an explicit compact solution, 
the generalized Vaidya solution (\ref{Vaidya}), 
can be given in a gauge in which one of the coordinates 
has been made to coincide with the dilaton field. 
We propose going a step further and  
taking a gauge such that each of the two coordinates coincides  
with each of the scalar fields, $\phi$ and $f$, in the theory. 
In other words, we will  
use $\phi$ and $f$ as coordinates. 

We are not going to dwell here on the quality 
of these coordinates, 
as they are, firstly and above all, a tool to help solve 
a certain system of equations. Therefore we are not going 
to analyze in this letter, for instance, how much of the spacetime 
can be covered with this system of coordinates. This and related  
questions will be considered in future communications. 
Nonetheless, these coordinates are quite natural 
-- hence the chosen name -- and incorporate much of what 
has been said in the literature about the necessity of using 
an internal time in gravity (see, for instance, Ref. 
\cite{[time],[observable]} and references therein). 
Our approach goes even further 
and uses not only an internal time,
but an internal space-time. The points of the 
space-time are described by physical quantities, 
$f$ and $\phi$, which are subject to dynamics. In fact, 
this approach may be regarded as a modelling  
(in two dimensions) of relationism. 

On the other hand, the technical 
advantages of these coordinates are apparent. 
Firstly, they render the equations in 
(\ref{eqofmotphi}) first order. 
Secondly, the arbitrary functions which appear, 
$V$ and $\Omega$, are functions of the coordinates, 
which should facilitate 
a generic treatment of all the models. 
Thirdly --and this is an unexpected fact,   
the origen of which remains hidden to us--,  
in these coordinates it turns out that 
 the equation of motion for the matter field, 
Eq. (\ref{eqofmotf}), follows 
from Eq. (\ref{eqofmotphi}). Therefore all the equations of 
motion are brought into a system of three first-order 
partial differential equations 
with two independent variables and three 
dependent variables (the component of the metric tensor). 

This system, if expressed in contravariant form  
\be 
\nabla^\mu\nabla^\nu\phi =
\frac12g^{\mu\nu}V -\frac{\Omega}2\left\{\nabla^\mu f\nabla^\nu f
-\frac12g^{\mu\nu}(\nabla f)^2\right\}\label{Tmunu2}\ee 
can be written in the form  

\be -\Gamma^{\phi,\mu\nu}=\frac12g^{\mu\nu}V-
\frac{\Omega}2\left\{\nabla^\mu f\nabla^\nu f
-\frac12g^{\mu\nu}(\nabla f)^2\right\}\label{Tmunu3}\ee
where $\Gamma^{\lambda,\mu\nu}$,  
the ``contravariant'' Christoffel symbols, involve 
the contravariant metric tensor only:  

\bea \Gamma^{\lambda,\mu\nu}
&\equiv&g^{\alpha\mu}g^{\beta\nu}\Gamma^\lambda_{\alpha\beta}\nonumber\\
&=&-\frac12\left(g^{\mu\rho}\partial_\rho g^{\lambda\nu}
+g^{\nu\rho}\partial_\rho g^{\lambda\mu}
-g^{\lambda\rho}\partial_\rho g^{\mu\nu}\right)\label{Gamma}\eea
It is clear that the system 
in Eq. (\ref{Tmunu3}) will take the 
simplest form if the three dependent variables 
are chosen to coincide with the components  
of the contravariant metric tensor $g^{\mu\nu}$. 

If we make $g^{\phi\phi}=F$, 
$g^{\phi f}=g^{f\phi}=G$ and $g^{f f}=H$,   
we have 

\bea \Gamma^{\phi,\phi\phi}&=&
-\frac12 F\p_\phi F-\frac12G\p_f F\label{1}\\
\Gamma^{\phi,\phi f}&=&-\frac12H\p_f F 
-\frac12 G\p_\phi F\label{2}\\ 
\Gamma^{\phi,ff}&=&-G\p_\phi G -H\p_f G 
+\frac12F\p_\phi H +\frac12 G\p_f H\label{3}\eea
and 

\bea T^{\phi\phi}&=&\frac{\Omega}2\left(G^2-\frac12FH\right)\label{11}\\
T^{\phi f}&=&\frac{\Omega}4GH\label{12}\\
T^{\phi\phi}&=&\frac{\Omega}4H^2\label{13}\eea 

Therefore, the equations of motion take the form 

\bea F\p_\phi F+G\p_f F=
FV-\Omega\left(G^2-\frac12FH\right)\label{21}\\
H\p_f F +G\p_\phi F=GV-\frac{\Omega}2GH\label{22}\\
G\p_\phi G +H\p_f G -\frac12F\p_\phi H -\frac12 G\p_f H=
\frac12HV-\frac{\Omega}4H^2\label{23}\eea 

Now, a bit of algebra with the first and second 
equations yields 

\be\left.\ba{l} \p_f F=-\Omega G\\
\p_\phi F = V +\frac{\Omega}2 H\ea\right\}\Rightarrow 
-\p_\phi(\Omega G)=\frac\Omega2\p_f H\label{eq1}\ee 
and 

\be H\p_f G -\frac12F\p_\phi H -G\p_f H =\frac12HV 
-\frac{\Omega}4H^2 +\frac{\p_\phi\Omega}\Omega G^2\label{eqlarga}\ee 
Hence, all the complexity of these models have been 
encapsulated in Eq. (\ref{eqlarga}), which in term of the 
single unknown function $D=F-J$  can be written 
(with obvious notation) in the form: 

\be -2D_{\phi}D_{ff} +2D_f D_{f\phi}
 +(D+J)(\Omega_\phi D_\phi -\Omega D_{\phi\phi}) 
-\Omega V D_\phi +\Omega (D_\phi)^2 - 
\frac{\Omega_\phi}{\Omega}(D_f)^2=0 \label{eqofmotofD}\ee

In summary, we have shown that in these coordinates the (contravariant) metric 
tensor $(g^{\mu\nu})$ can be expressed in term of a single function $D$:  

\be (g^{\mu\nu})=\left(\ba{cc}D+J&-\frac1\Omega D_f\\
-\frac1\Omega D_f&\frac2\Omega D_\phi\ea\right)
\label{contrametric}\ee 
This function $D$ is not arbitrary 
but has to obey Eq. (\ref{eqofmotofD}), which, therefore, encapsulates all 
the dynamical content of these models. 
We hope that the compactness and (relative) simplicity 
of this result will be useful to solve and understand 
two- and  higher-dimensional classical and quantum gravity. 

To finish, let us show with an example how this equation 
may actually serve to find solutions and perhaps  
the general trajectories of these models. If 
$\Omega=1$ (minimally coupled matter), 
then $D_\phi=0$ is a solution to Eq. (\ref{eqofmotofD}). 
Eq. (\ref{eq1}) implies then 

\be (g^{\mu\nu})=\left(\ba{cc}J+K&-\p_f K\\-\p_fK&0\ea\right)
\label{naturalVaidya}\ee 
with $K$ an arbitrary function of $f$, $K=K(f)$. 
Clearly, this is the Vaidya solution,  
which can be checked with 
the corresponding change of coordinates.  

\section*{Acknowledgements}
The author is grateful to the Spanish MEC, 
CSIC and IMAFF (Madrid) for a research contract. The author would like 
to thank J. Navarro-Salas for his very useful comments and suggestions. 

This work was partially supported by the 
Comisi\'on Interministerial de Ciencia y Tecnolog\'{\i}a 
and DGICYT. 

\end{document}